\documentclass[fleqn,usenatbib]{mnras}

\usepackage{newtxtext,newtxmath}

\usepackage[T1]{fontenc}

\DeclareRobustCommand{\VAN}[3]{#2}
\let\VANthebibliography\thebibliography
\def\thebibliography{\DeclareRobustCommand{\VAN}[3]{##3}\VANthebibliography}
\usepackage{soul}
\usepackage{graphicx}
\usepackage{amsmath}

\title[RRab lookalikes at different periods]{
RRab variables with identical light-curve shapes at different pulsation periods}
\author[J. Jurcsik et al.]{
Johanna Jurcsik,$^{1}$\thanks{E-mail: johanna.jurcsik@gmail.com}
\'Aron Juh\'asz$^{1,2}$
\\
$^{1}$Konkoly Observatory, ELKH CSFK, H-1121, Budapest, Konkoly Thege Mikl\'os \'ut 15-17., Hungary\\
$^{2}$E\"otv\"os Lor\'and University, H-1117, Budapest, P\'azm\'any P\'eter s\'et\'any 1/a, Hungary
}

\date{Accepted 2022 September 20. Received 2022 September 12; in original form 2022 July 19}

\pubyear{2022}

\begin{document}
\label{firstpage}
\pagerange{\pageref{firstpage}--\pageref{lastpage}}
\maketitle

\begin{abstract}
In this paper, we report on the detection of RRab stars with quasi-identical-shape light curves but period differences as large as $0.05-0.21$ d using the Galactic bulge data of the OGLE-IV survey. We have examined stars with shorter periods than the Oosterhoff~I ridge of the bulge. These stars generally have smaller amplitudes and larger Fourier phase-differences than the typical bulge RRab stars have at the same period. Many of these "anomalous" stars have good-quality light curves without any sign of the Blazhko modulation. Examining their Fourier parameters revealed that several of these stars show very similar light-curve to the typical bulge RR Lyrae. We found hundreds of quasi-identical-shape light-curve pairs with different periods between the "anomalous"- and the "normal"-position RRab stars based on the OGLE $I$-band data. The OGLE $V$-band, and the archive VVV and MACHO surveys $K_s$-, $b$- and $r$-band data of these stars were also checked for light-curve-shape similarity. Finally,  149 pairs with identical-shape light curves in each available photometric band were identified.
Calculating the physical properties of the variables using empirical formulae, on average, $-0.5$~dex, $-0.13$~mag, 0.67, and 165~K differences between the [Fe/H], $M_V$, $R/R_\odot$, and $T_{\mathrm{eff}}$ values of the members of the pairs are derived, being the short-period stars less metal-poor, fainter,  smaller and  hotter than the long-period variables. To explain the existence of variables with different physical properties and pulsation periods but with identical-shape light curves is a challenging task for modelling.
\end{abstract}


\section{Introduction}

Pulsating stars have always been central among astronomy targets, 
because their global physical parameters (mass, luminosity, effective temperature, chemical composition) and their internal structures determine the emitted and detectable light and light variations. Accordingly, most of their physical properties are deducible from different-band photometric time-series data, which are available today in large amounts thanks to the continuous development of the photometric techniques.

The  periods and the light-curve amplitudes/shapes of RR Lyrae (RRL) stars depend on the physical properties of the stars as it was  already shown by the early radiative and convective model results \citep{Christy, BS}. 
N. Simon and co-workers were the very first who found relations between the observed $\varphi_{31}$ Fourier parameter of the light curve and the physical properties of first-overtone RRL stars using hydrodynamic models \citep[see e.g.,][]{Simonclem}.
The frequency dependent radiative transfer calculations of nonlinear convective models published by \cite{Dorfi} showed in detail how the Fourier parameters of the light curve vary with varying physical properties of the stars in different photometric bands.

Among the observable parameters of the light variation of classical pulsating stars like RRLs and Cepheids, the pulsation period is the determinant one. Pulsation model calculations show how the pulsation period depends on the fundamental physical parameters of the variables \citep[see e.g.,][]{Marconi07}, and restrict the possible domains of the physical parameters of the stars.
The period identifies the pulsation mode in most cases; in addition, using the period-luminosity (P-L) and the period-luminosity-colour (P-L-C) relations it is the primary indicator of the luminosity, and, as a consequence, of the distance.

The shape of the light curve also holds information on the basic specifics (chemical composition, luminosity, effective temperature, stellar radius, surface gravity) of the variables. In eliciting this information, both hydrodynamic modelling and empirical methods are successful.

 However, after the pioneering work of \citep{SimonDavis}, very few attempts were made to model the light variations in different photometric bands with sufficient accuracy \citep[see e.g.,][]{Dorfi,Marconi05,Das}. Recently, the fundamental parameters of Cepheids and RRL stars in the Magellanic Clouds and the Galactic bulge were derived using nonlinear 1D convective models by \cite{Bellinger}. Nevertheless, the fine structure of the light curve was not considered in this study. Some basic light-curve parameters, viz. period, skewness, acuteness and amplitudes were actually used to match the observations to the model predictions using the artificial intelligence neural network training method.

The empirical relations utilising the period and light-curve-shape parameters of RRL stars (e.g., Fourier amplitudes and epoch independent phase differences) yield estimates on the [Fe/H], the absolute $V$ magnitude ($M_V$), and the effective temperature ($T_{\mathrm{eff}}$) with similar accuracy as spectroscopic studies \citep[e.g.,][] {jk96,Jur98,Kw01,Nemec13,Dekany21}. The existence of these relations is a natural consequence of the fact that the surface gravity and effective temperature variations along the pulsation cycle, which affect the radiative and convective energy transport, depend on the global physical properties of the star.

RR Lyrae stars are supposed to constitute mono-metallic samples in most of the globular clusters. The Fourier parameters of their light curves show progressions with increasing pulsation periods (see e.g. \citealp{Bram11} for M72 and \citealp{Jur17} for M3). In the case of Cepheids, a continuous change of the light-curve shape (Hertzsprung-progression) is also detected \citep[see e.g.][]{Kovacs89}. 

Based on these observations and model predictions, one would assume that the light-curve shapes of RRL stars, neither of similar nor of different metallicities, are completely identical at different pulsation periods. 
Although the existence of RRL stars with identical-shape light curves at different pulsation periods has not been proved or refused either theoretically or by observations, the listed direct and indirect information seems to contradict their existence.

Several plots of different Fourier-parameter combinations of the Galactic bulge RRab stars (fundamental-mode RRLs) were shown in the papers of \citet[][hereafter P17]{Prudil17} and \cite{Prudil19a}. Looking at these Fourier parameters versus period plots, there are several stars in the sample that look to be outliers. These stars are usually below the Oosterhoff I (OoI) ridge in the period-amplitude plot. They also appear at shorter periods, on the left side of the OoI ridge, on the period versus Fourier phase differences, and on period versus amplitude-ratio diagrams. Most of these stars were identified by P17 to show the Blazhko effect, which distorts the light curve, but there are hundreds of such stars among the bona fide non-Blazko RRab variables selected by P17. 

Possible explanations for the anomalous positions of these stars are as follows:
\begin{itemize}
    \item Blazhko modulation, which was hidden in the shorter database analysed by P17,
    \item blending or some other photometric defect,
    \item significantly different physical parameters (e.g., [Fe/H]) of these stars as for the bulk of the bulge RRLs.
\end{itemize}

In order to find some special cases showing the temporal appearance of the Blazhko phenomenon \cite[see e.g. V114/M3 and OGLE-BLG-RRLYR-07605 ][]{Jur13,Sosz14}, we investigated these outlying-position stars. In several cases, the photometry seems to be affected by crowding, and the light curve is very noisy. However, many of these stars have typical RRL-like shapes and good-quality light curves with no indication of any modulation or photometric defect. The positions of these stars on the plots showing different combinations of the Fourier parameters but the period do not indicate any systematically anomalous behaviour. They fit the overall patterns traced out by the normal OoI/OoII RRLs of the bulge. 
Consequently, these stars seem to be ordinary RRL stars of probable different physical properties than most RRL stars in the bulge.

As the shape of the light variation depends on the physical properties of the star, it may be assumed that the light curves of these stars have to be somewhat different from the light curves of the normal bulge sample stars.
However, comparing the light curves of the "anomalous"-position and the normal bulge RRab stars we noticed that their shapes are identical within the uncertainty limits in several cases. Only the periods of these stars differ noticeably.

In the present paper, we aim to show that RRab stars with quasi-identical light-curve shapes do exist at significantly different pulsation periods.

\section{Data and method}

The fourth survey of the Optical Gravitational Lensing Experiment (OGLE IV) provides light curves of over 38000 RRL stars in and towards the Galactic bulge \citep{Sosz14}. As the basis of the present work, we used the updated OGLE IV catalogue, which includes observations until the 2017 observing season. 

P17 identified 3341 variables showing Blazhko modulation using a selected sample of variables from the 2014 issue of the OGLE IV catalogue. Omitting these Blazhko stars, the uncertain Blazhko candidates and also the period-changing ones from the P17 catalogue, and applying the criteria given in E.qs. 1.-4. in \cite{Prudil19a} to exclude possible foreground stars, we remained at 4862 bona fide non-Blazhko stars of the Galactic bulge RRab sample. This is the primary data set for our study.
\cite{Prudil19a}, studying the Oosterhoff properties of the bulge RRL stars, defined a similar sample, but they considered only variables with observation in the Vista Variables in the Via Lactea (VVV) survey as well, and this was not a criterion in our selection.
However, the OGLE $V$, the Macho instrumental $b$ and $r$ \citep{Alcock} and the VVV $K_s$-band \citep{Minniti} data, if available, were also used for the stars in our interest. 

The amplitudes of the public VVV $K_s$-band light curves derived from different-size aperture photometry \citep{Dekany18,Hajdu20} are slightly different, therefore, there is some ambiguity in the amplitudes of the $K_s$ light curves. Typically, we used $K_s$-band photometry showing light curves with the smallest residual scatter.

Hereafter, the RRL stars from the bulge OGLE sample (OGLE-BLG-RRLYR-NNNNN) are referred to as their number, NNNNN, alone. 

Period04 \citep{LenzBerger} software was used for frequency analysis to generate Fourier parameters and their errors.

\begin{figure*}
\begin{center}
\includegraphics[width=1.0\textwidth]{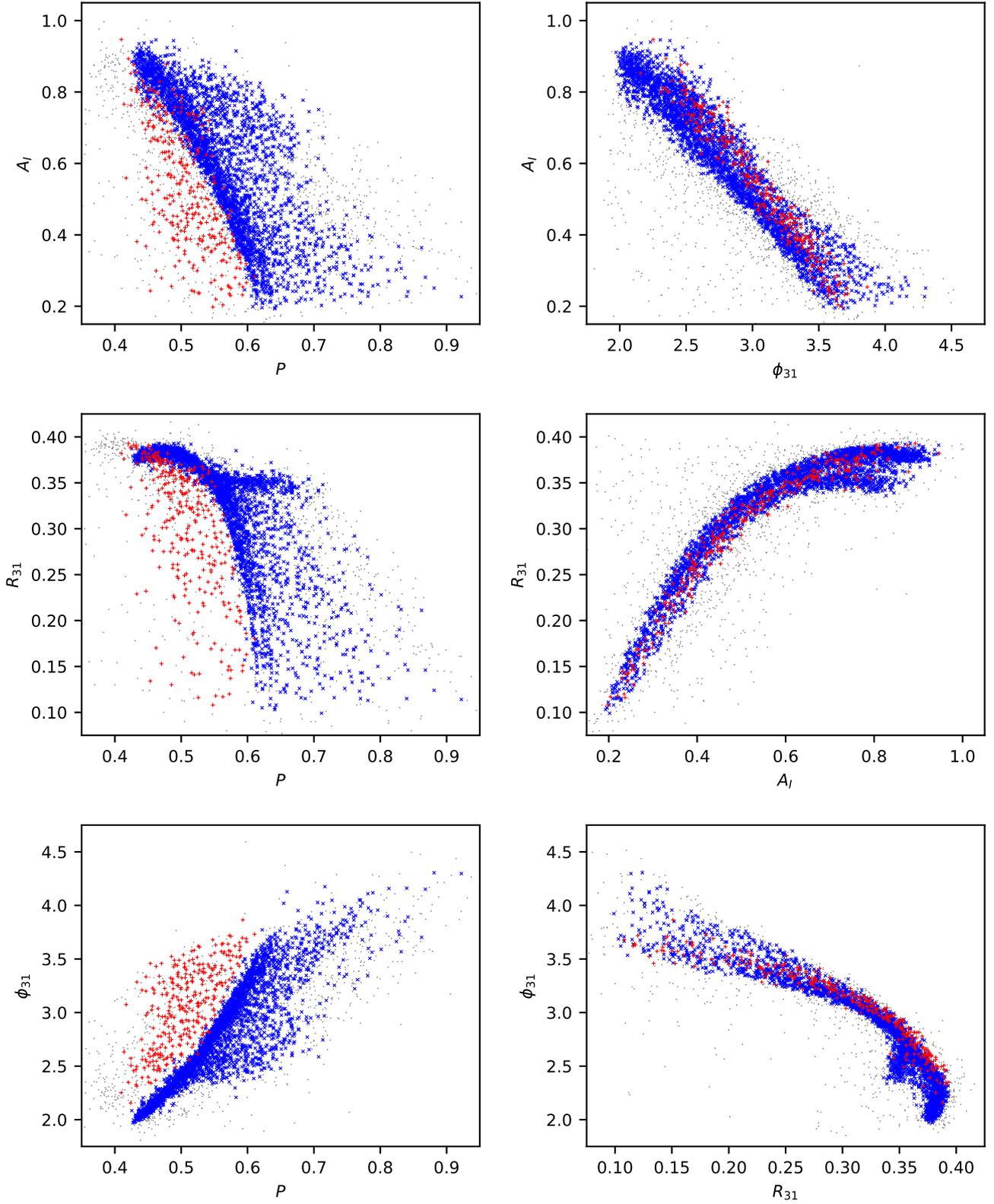}

\caption{Light-curve-parameter spaces of the OGLE $I$ data of the Galactic bulge non-Blazhko RRab stars are shown in the panels. $A_I$, $R_{31}$ and $\phi_{31}$ versus the period are shown in the left-hand panels. The right-hand panels display the relations between the $A_I$, $R_{31}$ and $\phi_{31}$ parameters. Grey dots, red crosses and blue x symbols denote the total sample of stars, the anomalous (A) and the normal (N) samples of RRab stars selected for the analysis, respectively.}
\label{fig:Fourier-period-plot}
\end{center}
\end{figure*}

\begin{figure*}
\begin{center}
\includegraphics[width=1.0\textwidth]{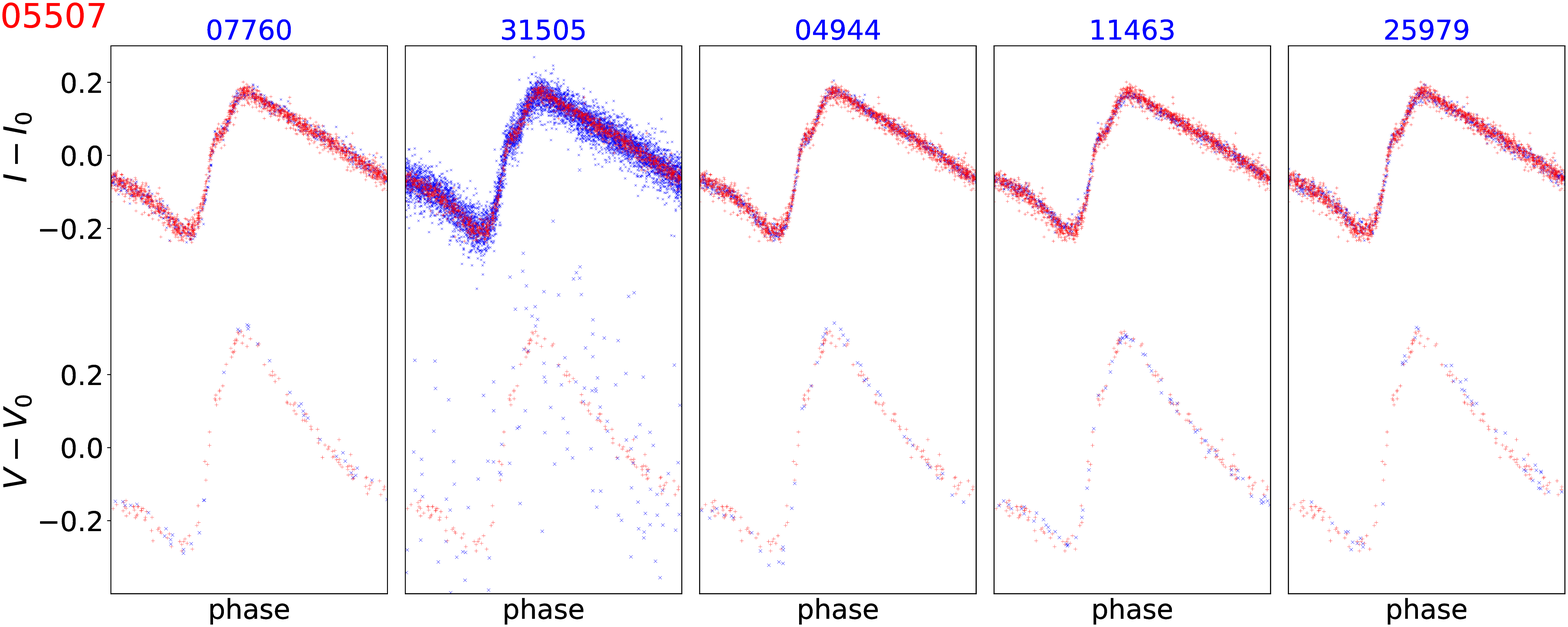}
\caption{One example of the pair-selection results is shown. The phase- and magnitude-matched $I$- and $V$-band light curves of the anomalous variable, 05507 (red crosses), are compared with the light curves of its five best normal pairs (blue x symbols) in the top and bottom panels, respectively. These five pairs are supposed to have similar light curves based on their $I$-band Fourier parameters, and the differences between their periods are larger than 0.05 d. The names of the selected normal stars are given at the top of each panel. 
Only the last two normal stars (11463 and 25979) are accepted as of probable identical shapes with 05507. The other three are rejected: the first and the third pairs suggest amplitude differences in the $V$-band, and the light curves of 31505 are too noisy.}
\label{fig:lc-examples-plot}
\end{center}
\end{figure*}

\section{Selection of the shorter-period, "anomalous", and the normal samples of RRab stars}

First, the sample of the "anomalous" RRab stars was selected based on their positions on the period - $A_{I}, R_{21}, R_{31}, \phi_{21}, \phi_{31}$ plots, using the data given in the OGLE database. Any star with light-curve parameters below the lower, left-side boundary defined by the main sample of RRab stars on any of these plots was regarded as "anomalous" compared to the bulk of the bulge variables.

As our aim was to find similar-shaped RRLs to the "anomalous" stars among the normal bulge variables, at second, variables that did not fit the global trends defined by the bulge sample in the plots of any combinations of the light-curve parameters without the period (i.e. $A_{I}, R_{21}, R_{31}, \phi_{21}, \phi_{31}$) were skipped. This restriction ensures that the light-curve shape of a star in the remaining sample does not differ significantly from the light curves of normal bulge RRLs. 

Variables with noisy $I$-band light curves were omitted.

The image charts provided by the OGLE team were visually inspected for any contamination of the neighbouring stars, but no apparent reason for any photometric defect was detected in any of the selected stars. However, we have to note that this process does not exclude the possibility of unresolved blends in some cases.

At the end, we arrived at a sample of 312 "anomalous" (A) stars, whose light curves were supposed to be similar to the light curves of other RRab stars in the bulge, but their periods were systematically shorter. 

To find light curves similar to these "anomalous" stars, the total sample of bulge RRLs are also cleaned from outliers. Variables with parameters not fitting the parameter spaces outlined by the bulk of the bulge sample are ignored. This process led to the selection of the normal (N) sample of 3068 RRab stars located at proper positions in any parameter space, including the period.

The total sample of 4862 non-Blazhko stars listed by P17, and the selected A and N samples of the variables are plotted by different symbols in the different light-curve-parameter plots shown in Fig.~\ref{fig:Fourier-period-plot}.

\section{Searching for identical-shape light curves in the two samples}

To this end, we searched for the smallest values of the square roots of the sums of the normalised quadratic differences between the $A_{I}, R_{21}, R_{31}, \phi_{21}, \phi_{31}$ parameters of the $I$-band light curves of the A- and the N-sample stars. 
As the light curves of variables with similar periods may also be very similar, the search was restricted to variables with period differences larger than 0.05 days.

For each variable of sample A, we selected the first five stars with the smallest differences from sample N.
The OGLE $I$- and $V$-band light curves of these pairs were then fitted in magnitude and phase, and their similarity was checked visually to find pairs of identical shapes, even in the smallest features (bumps, humps, etc.) of the light curves.

An example of this process is shown in Fig.~\ref{fig:lc-examples-plot}. The $I$ and $V$ band light curves of the five best pairs are compared for the 05507 "anomalous" RRL. The normalised quadratic differences between the parameters of the pairs increase from left to right. Although it is somewhat subjective which pairs are selected as candidates for identical respect to their light-curve shapes, we tried to be as rigorous as possible and skipped every pair with any small differences in their light curves. 
In the case of 05507, the agreement between the phase- and magnitude-adjusted light curves seems to be satisfactory for two of its five best pairs.

\begin{figure*}
\begin{center}
\includegraphics[width=1.0\textwidth]{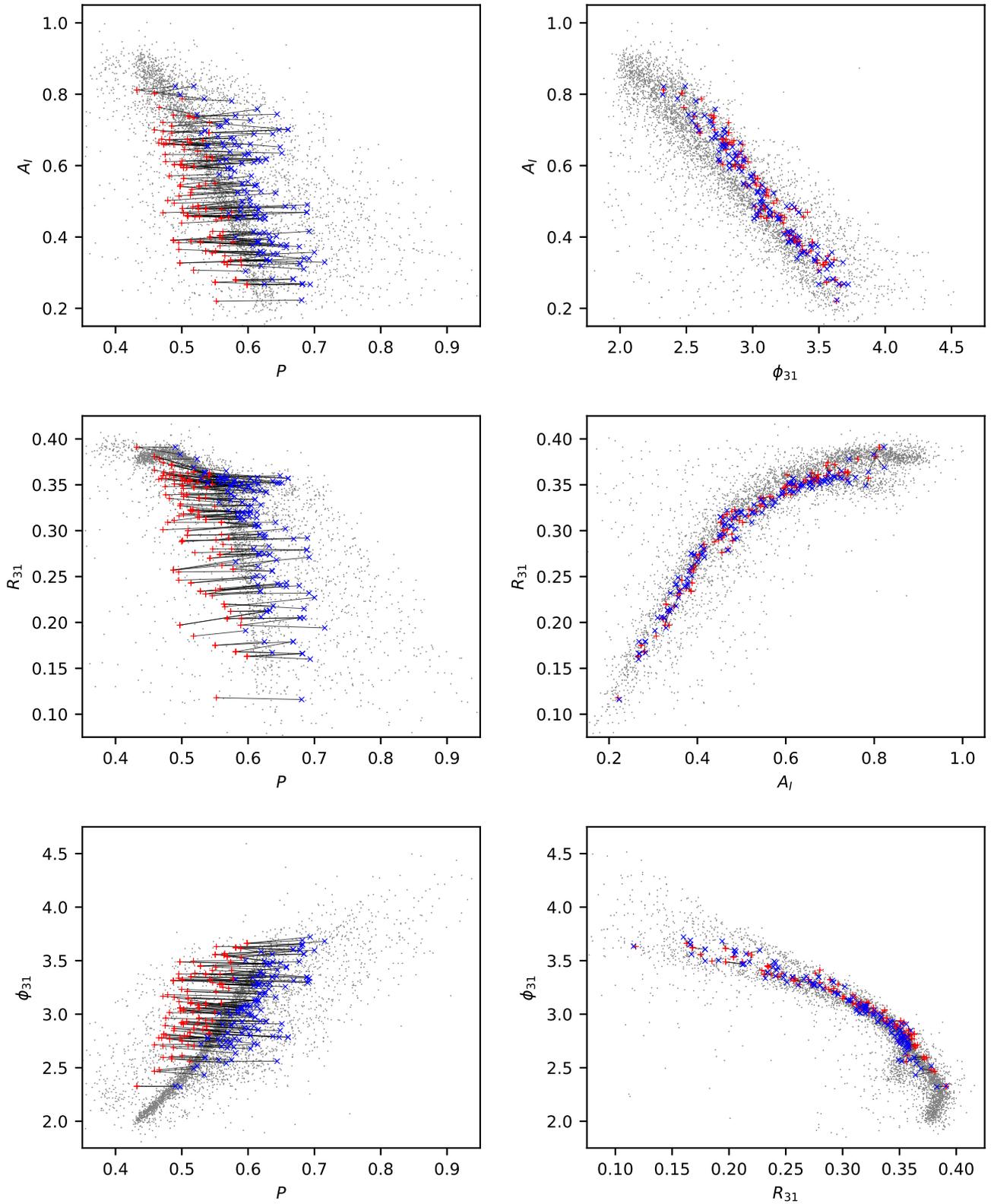}

\caption{The same plots as in Fig.~\ref{fig:Fourier-period-plot} are shown for the 95 anomalous (red crosses) RRab stars and their 116 normal pairs (blue crosses) in the figure. The pairs are connected by grey lines. The period differences of the A (short-period) and the N (long-period) members of the pairs are larger than 0.05 d, but the light curves of their available observations do not indicate any differences in their shapes.}
\label{fig:pairs-plot}
\end{center}
\end{figure*}

\begin{figure*}
\begin{center}
\includegraphics[width=1.0\textwidth]{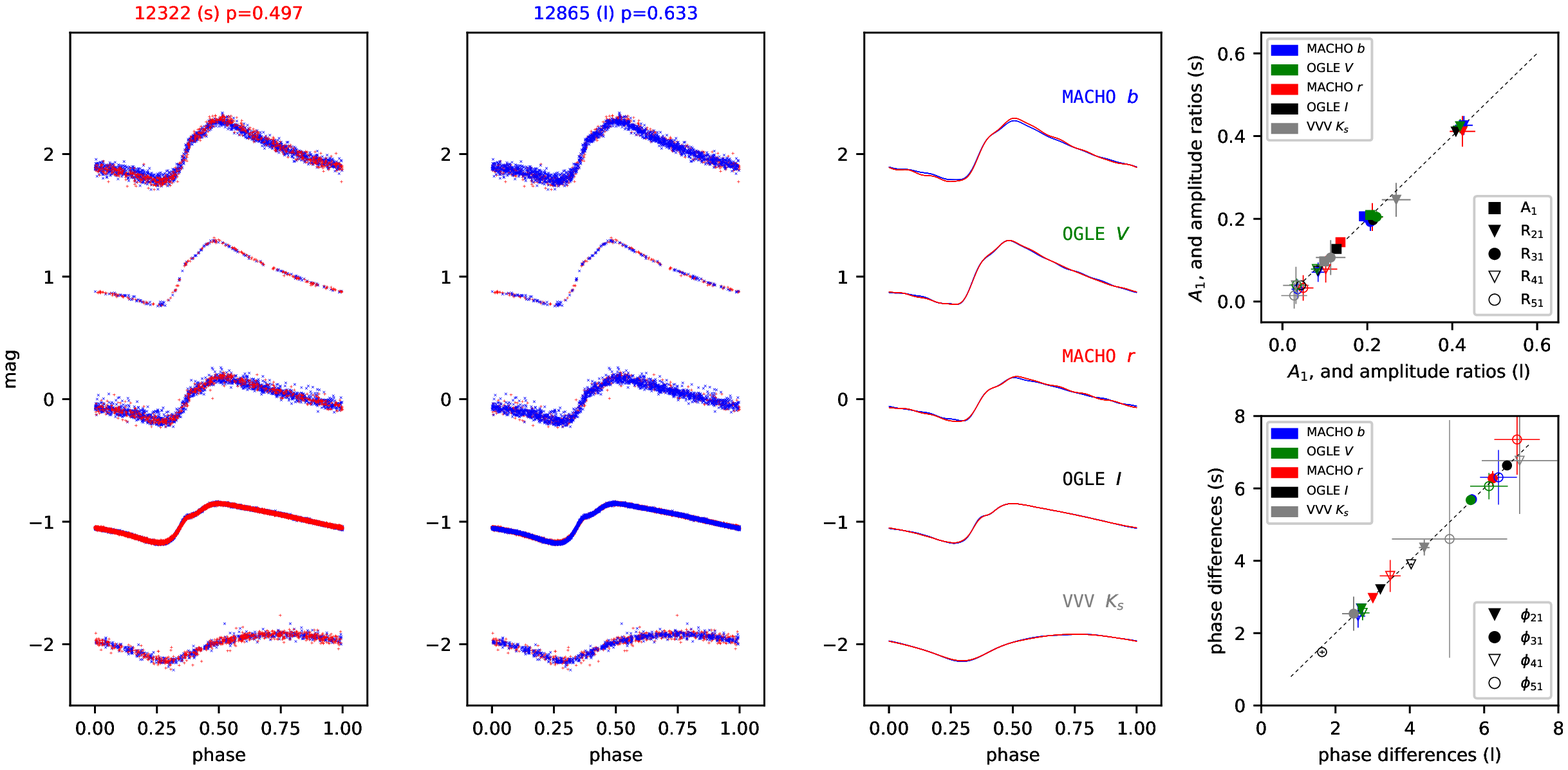}
\includegraphics[width=1.0\textwidth]{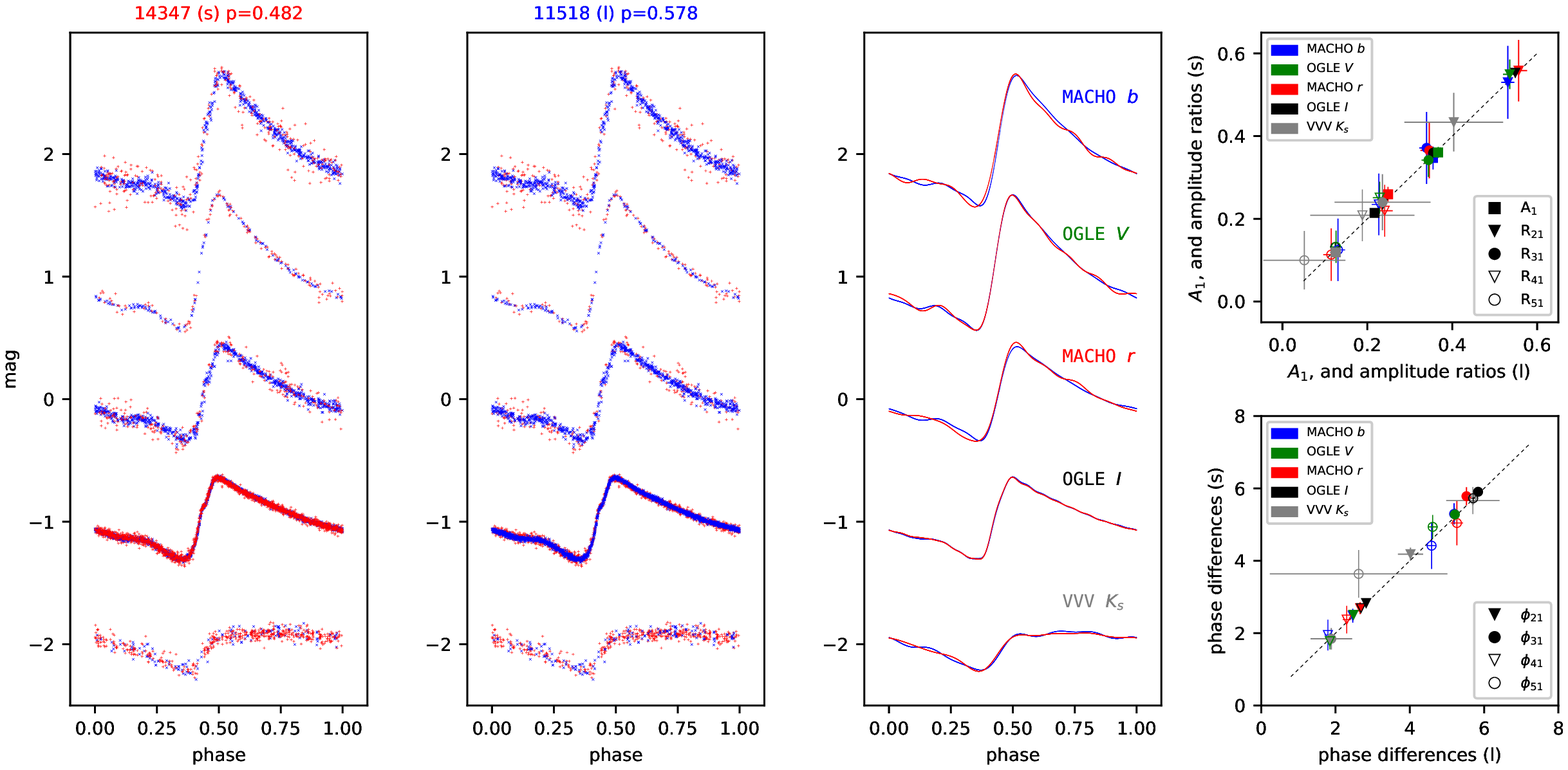}

\caption{Two examples for the quasi identical-shape light curves at different pulsation periods are shown.
The top and bottom panels display the light curves and the Fourier solution results for the 12322-12865 and the 14347-11518 pairs, respectively.
The OGLE number and the pulsation period of the stars are given at the top of the light-curve panels.
The light curves of the short- and the long-period members of the pairs are over-plotted on each other by red crosses and blue x symbols in different order in the two left-side panels. The phase matched, mean-magnitude corrected values (arbitrarily shifted) are shown. The fitted light curves are drawn in the third panels. The right-hand panels compare the Fourier parameters of the two stars. The $3 \sigma$ formal errors of the parameters and the equality lines are also drawn in in these plots.} 
\label{fig:12322}
\end{center}
\end{figure*}

Notwithstanding that the $V$ and $I$ band OGLE light curves of the pairs selected at this step do not show any noticeable differences, because of their different periods, we have guessed that at shorter or longer wavelengths some systematic differences may occur.
Therefore, the MACHO and the VVV data, if they are available and are of good enough quality are also checked for similarity. Unfortunately, the MACHO data of most of the stars in question are too noisy to draw any conclusion in these bands, but differences between the $K_s$-band light curves of several pairs are indeed detected.

The MACHO $b$-band is somewhat bluer than the photometric  band, and the MACHO $r$-band is between the $V$ and $I$ bands \citep{Alcock}. Therefore, slight differences between the $b$-band light curves of variables with identical $I$ and $V$ light-curve shapes may occur, but differences in their $r$-band light curves are not supposed to be detected. However, there are some pairs that do not show any differences in the $V$- and $I$-bands but both the MACHO $b$ and $r$ light curves of one or both members of the pair have different shape and/or amplitude. Some photometric defects and the increased scatter of the MACHO light curves mimicking some differences in the shapes of the light curves may be the reason for this behaviour. The MACHO observations were obtained about twenty years earlier than the OGLE IV measurements, and the multi-mode and the modulation properties of RRL stars may change abruptly on a timescale of years resulting in a change in the shape and or the amplitude of the mean light curve \cite[see e.g.,][]{Sosz14}. Therefore, the possibility that the character of the light curve of one of the stars has changed between the MACHO and OGLE observations cannot be excluded either. 
Whatever the reason for the contradicting behaviour of the MACHO light curves of these pairs, they were also removed from our final sample of pairs.

Finally, we remained at 149 quasi-identical shape, but different-period pairs between 95 and 116 RRLs of the A and N samples. Good accuracy VVV and MACHO data are available for 106 and 11 of these pairs, respectively.
The positions of these pairs are shown and connected by lines in Fig.~\ref{fig:pairs-plot}. This figure displays the same plots as shown in Fig.~\ref{fig:Fourier-period-plot}.
The short- (A) and long-period (N) members of the pairs are denoted by the "s" and "l" subscripts in the following.

Fig.~\ref{fig:12322} shows examples of the similarity of the light curves of the pairs. The light curves of the short-period and the long-period members of the pairs are over-plotted on each other in different order in the first two panels, and the third panels show the synthetic light curves of the stars based on appropriate-order Fourier fits to the data. The orders of the fits are typically 10, 10, 10, 16 and 5 in the $b$-, $V$-, $r$-, $I$,- and $K_s$-bands, respectively, but higher or lower orders are used in some cases if necessary. The right-hand panels of the figures compare the low-order Fourier parameters of the two stars. 

The period differences of the two pairs shown in Fig.~\ref{fig:12322} (12322-12865 and 14347-11518) are 0.13 d and 0.10 d, but no sign of any differences in their light curves are observed in any of the photometric bands shown. 

\begin{table}
\label{tab:1}
	\centering    
	\caption{The identification of the 149 pairs of RRab stars with quasi identical-shape light curves at different periods and their photometric data used. The full table is available as a supplementary material to this paper.}

\begin{tabular}{llllrrrrr}

\hline 
ID$_{\mathrm{s}}$&Perid$_{\mathrm{s}}$&ID$_{\mathrm{l}}$&Period$_{\mathrm{l}}$&$b$&$V$&$r$&$I$&$K_s$\\
\hline
00234   &	0.56378   &	02039	&	0.68431	&-&+&-&+&-	\\
00234   &	0.56378   & 13632	&	0.63782	&-&+&-&+&-	\\
00265	&	0.57375   &	08810	&	0.62815	&-&+&-&+&+	\\
00265	&	0.57375   &	12865	&	0.63272	&-&+&-&+&+	\\
00268   &	0.47287   &	08672	&	0.53916	&-&+&-&+&-	\\	
\multicolumn{9}{l}{......}\\
\hline
		\end{tabular}
\end{table}

Table~\ref{tab:1} lists the identification of the 149 pairs detected to have similar-shape light curves at different periods and the photometric bands of their observations utilized.

In the course of the selection process of the pairs, we also detected several pairs with light curves showing minor differences in amplitude and/or shape and yet their low-order Fourier components do not show noticeable differences.

\begin{figure*}
\begin{center}
\includegraphics[width=1.0\textwidth]{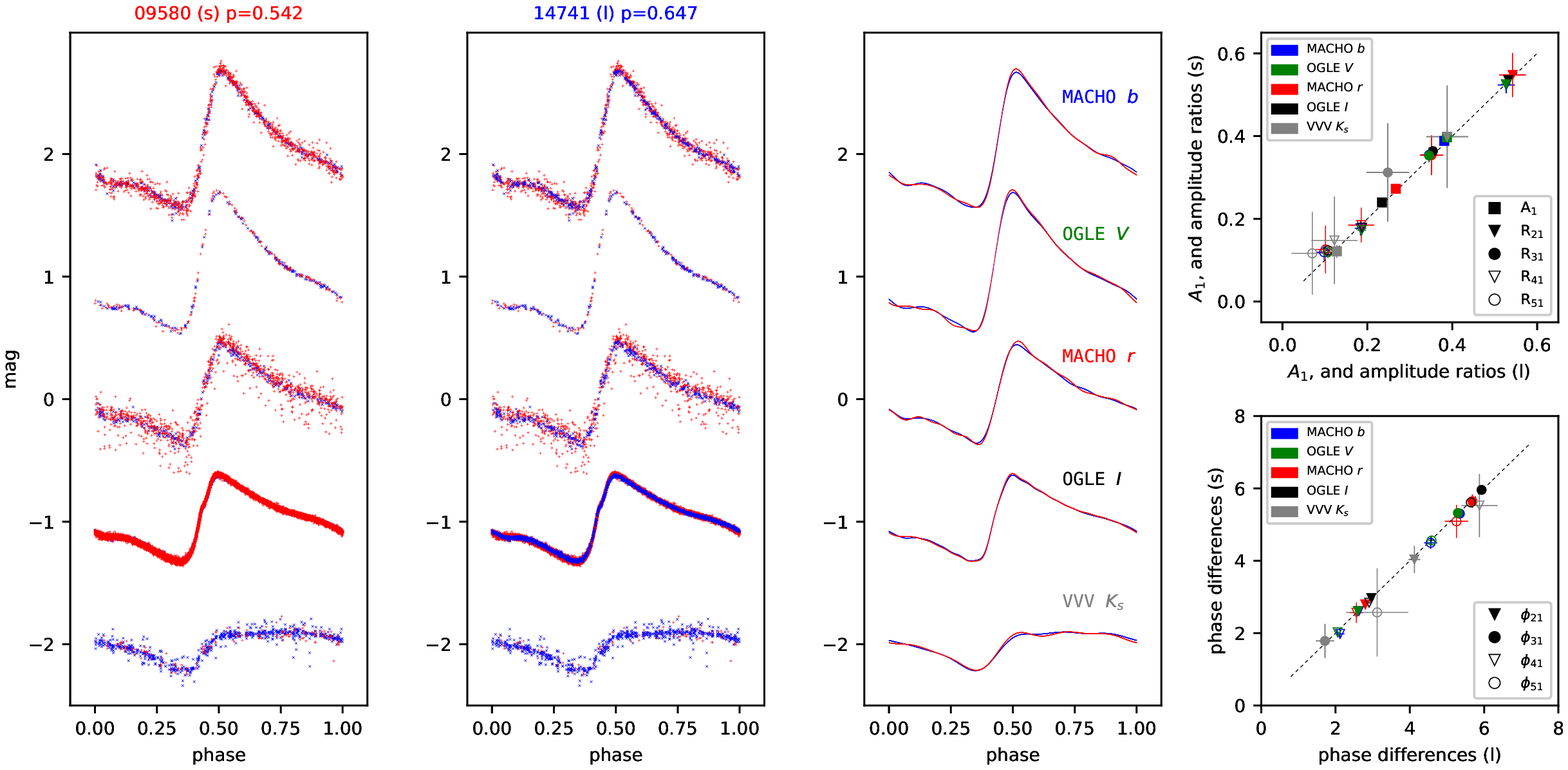}
\includegraphics[width=1.0\textwidth]{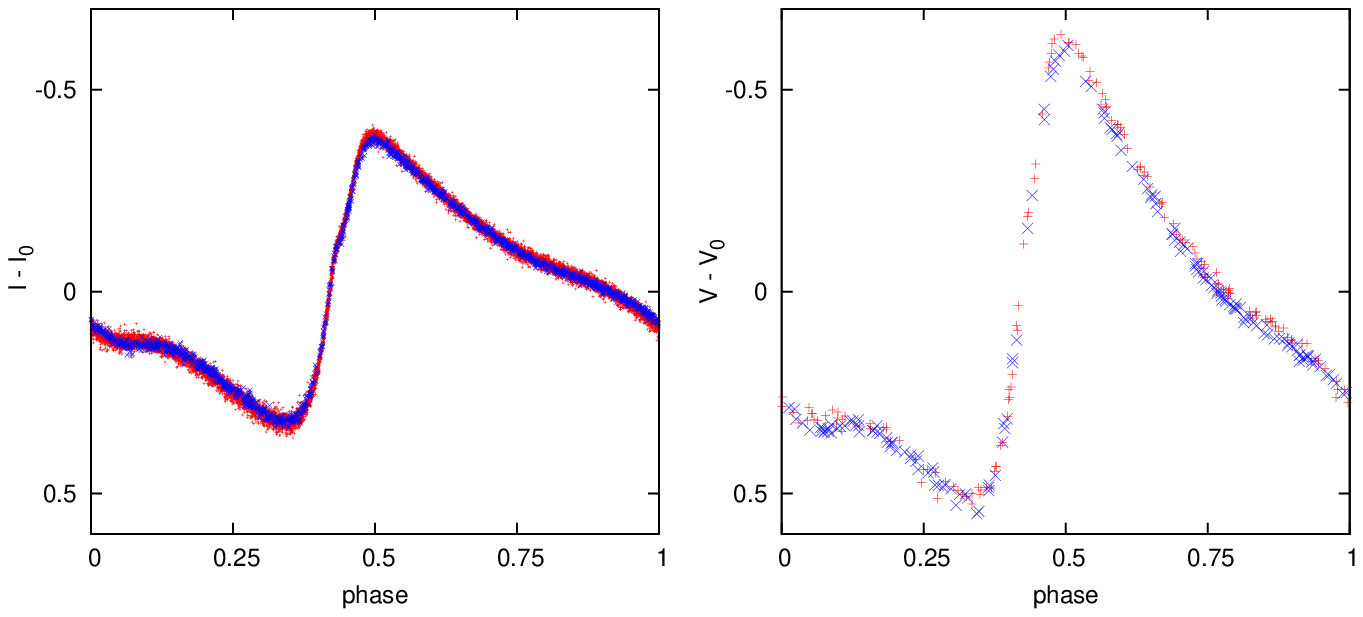}

\caption{An example of very small differences between the light curves of different-period stars. The top panels display the same plots as shown in Fig.~\ref{fig:12322}, and the bottom panels display the OGLE $I$ and $V$ data of the two stars on larger plots. The Fourier parameters of the pair do not indicate any significant differences, however, the light curves are slightly different in amplitude and in the shape of the bump on the descending branch.}
\label{fig:09580}
\end{center}
\end{figure*}

An example is shown in Fig.~\ref{fig:09580}, which displays similar plots as shown in Fig.~\ref{fig:12322}. Nonetheless that the low-order Fourier components of the two stars are very similar (right-hand panels in the top figure), a close inspection of the light curves (bottom panels) reveals small but characteristic differences e.g., between the shapes of the bumps and in the amplitudes. Actually, a normal pair (or pairs) with such small differences is/are found nearly to all the RRLs of the A sample. 

As the automation-based algorithms working on mass photometry data use typically limited numbers of parameters, several such quasi-twin pairs of different period variables occurring at any period are supposed to be detected in these databases.

\section{Physical parameters of the pairs}\label{sect:phys}

The light-curve parameters are good indicators of the fundamental physical properties of RRL stars. 

The physical parameters ([Fe/H], $M_V, R/R_\odot, T_{\mathrm{eff}}$) of the stars are estimated using the formulae given in \cite{Smo05,Dekany21, Kw01,Marconi15} and \cite{Jur98}, respectively. To calculate [Fe/H] both the \cite{Smo05} formula transformed to the UVES metallicity scale of globular clusters as given in \cite{Jur21,Jur22} and the formula given by \cite{Dekany21} recently, are applied. As a consequence, two values for the $R$ and $T_{\mathrm{eff}}$ values are derived depending on which metallicities are used.
Following \cite{ArellF10} 0.41 mag is taken for the zero point of the \cite{Kw01} $M_V$ formula.
To convert [Fe/H] to $Z$, when calculating the radius, the [$\alpha$/Fe] elemental ratio was taken to be 0.3 based on the bulge sample of metal-poor red giants and clump stars ([Fe/H]$<-0.5$) published by \cite{Duong}. This $\alpha$-element enhancement is somewhat larger than obtained for field RRL stars by \cite{Crestani21} ([$\alpha/{\mathrm{Fe}}]=0.2$ at [Fe/H]$=-1.0$) recently, most probably because of the different chemical evolution of the galactic disk and the bulge. Note also that, actually, the  $\alpha$-element enhancement  also depends on the [Fe/H], however, this dependency has no significant effect on the results shown here. The results obtained using the \cite{Dekany21} [Fe/H] formula are denoted by an asterisk hereafter.

Table~\ref{tab:2} lists, the derived [Fe/H], $M_V, R/R_\odot, T_{\mathrm{eff}}$ values of the short- and long-period members of the pairs.

As the pulsation period is the most dominant parameter in each of the relations and that there are no significant differences between any other Fourier parameters of the pairs, it is not surprising that systematic differences between the calculated physical parameters of the members of the pairs are detected.

The average differences between the parameters  ($Param_{\mathrm{l}}-Param_{\mathrm{s}}$) of the 149 pairs are: \newline
$\overline{\Delta{P}}=0.09$ d, s.d. 0.03;\newline
$\overline{\Delta{\mathrm{[Fe/H]}}}=-0.42$, s.d. 0.14; \newline  
$\overline{\Delta{\mathrm{[Fe/H]}}^*}=-0.59$, s.d. 0.20; \newline$
\overline{\Delta{M_V}}=-0.13$ mag, s.d. 0.04; \newline
$\overline{\Delta{R/R_\odot}}=0.63$, s.d. 0.21;\newline 
$\overline{\Delta{R/R_\odot}^*}=0.72$, s.d. 0.24;\newline
$\overline{\Delta{T_{\mathrm{eff}}}}=-169$ K, s.d. 66; \newline
$\overline{\Delta{T_{\mathrm{eff}}}^*}=-162$ K, s.d. 64.

Supposing that the spread in the masses of RRL stars in the sample does not exceed 0.02 $M_\odot$, as a result of  the systematically   larger  radius of the long-period stars, their surface gravity  is expected to be about 0.1 dex smaller than the log$g$ of their short-period counterparts.

Being the long-period member of the pairs  the brighter,  the larger and the smaller surface gravity star,  the possibility that they are more evolved horizontal-branch stars cannot be excluded.  However, the $R$, $M_V$ and log$g$ differences are partly/mostly the consequences of the chemical composition and period differences between the two samples. Therefore, the differences in the  evolutionary status of the variables may not be indeed significant. 

The difference between the observed mean $V$ and $I$ magnitudes of the two samples are $-0.21$ mag and $-0.24$ mag, being the short-period stars 0.2 mag fainter than their long-period counterparts, on average. This is in relatively good agreement with the calculated $-0.13$ mag mean $M_V$ differences of the pairs. However, taking into account the large and heterogeneous reddening of the bulge, and its extension, this agreement might just be accidental. Despite this, the agreement between the observed and the measured brightness differences between the two samples indicates that there is no systematic difference between the locations of the two samples, i.e., both the short- and the long-period members of the pairs belong to the galactic Bulge population of RRL stars.

\cite{Marconi18} investigating the impact of the He content on the absolute magnitudes in different bands and on the period of RRL stars, derived a formula that connects the mean $M_V$ of RRLs to their [Fe/H] and $Y$ contents. Using the $M_V$ and the [Fe/H] values of the stars to estimate $Y$, the mean differences between the $Y$ values of the short- and the long-period members of the pairs are 0.000 and 0.006 depending on which $I$-band metallicity formula are used. The He content of the more metal-poor, long-period sample is marginally, if at all, larger than those of the short-period sample.
The estimated mean $Y$ values of the short and the long-period samples are 0.207/0.208 and 0.211/0.217 using the \cite{Dekany21} and the \cite{Smo05} [Fe/H] formula, respectively.
Taking into account the uncertainties of the method the agreement with the canonical value of $Y=0.24$ of RRL stars is satisfactory.

\begin{table*}
\label{tab:2}
	\centering    
	\caption{Physical parameters of the pairs derived from the Fourier parameters of the light curves. Results using the \citet{Dekany21} $I$-band [Fe/H] formula are denoted by an asterisk. The full table is available as a supplementary material to this paper.}
		\begin{tabular}{c@{\hspace{4pt}}c@{\hspace{6pt}}c@{\hspace{4pt}}c@{\hspace{6pt}}c@{\hspace{2pt}}c@{\hspace{2pt}}c@{\hspace{2pt}}c@{\hspace{6pt}}c@{\hspace{2pt}}c@{\hspace{2pt}}c@{\hspace{2pt}}c@{\hspace{5pt}}c@{\hspace{5pt}}c@{\hspace{10pt}}c@{\hspace{5pt}}c@{\hspace{5pt}}c@{\hspace{5pt}}c}
		\hline 
ID$_{\mathrm{s}}$&Per$_{\mathrm{s}}$&ID$_{\mathrm{l}}$&Per$_{\mathrm{l}}$&
${\mathrm{[Fe/H]}}_{\mathrm{s}}$&
${{\mathrm{[Fe/H]}}_{\mathrm{s}}^*}$&
${\mathrm{[Fe/H]}}_{\mathrm{l}}$&
${{\mathrm{[Fe/H]}}_{\mathrm{l}}^*}$&
${R/R_\odot}_{\mathrm{s}}$&
${{R/R_\odot}_{\mathrm{s}}^*}$&
${R/R_\odot}_{\mathrm{l}}$&
${{R/R_\odot}_{\mathrm{l}}^*}$&
$M_{V_{\mathrm{s}}}$&
$M_{V_{\mathrm{l}}}$&
$T_{{\mathrm{eff}_\mathrm{s}}}$&
${T_{{\mathrm{eff}_\mathrm{s}}}}^*$&
$T_{{\mathrm{eff}_\mathrm{l}}}$&
${T_{{\mathrm{eff}_\mathrm{l}}}}^*$\\

 \hline
00234	&	0.56378	&	02039	&	0.68431	& $-0.70$& $-0.75$	& $-1.22$&	$-1.48$	&5.17&5.19&5.99&6.13&	0.72&0.56	&	6502	&	6502& 6325 & 6337	\\
00234   &	0.56378 &	13632	&	0.63782	& $-0.70$& $-0.75$	& $-1.07$&	$-1.26$	&5.17&5.19&5.70&5.80&	0.72&0.62	&	6502	&	6502& 6404 & 6413	\\
00265	&	0.57375	&	08810	&	0.62815	& $-0.84$& $-0.91$	& $-1.07$&	$-1.24$	&5.28&5.32&5.65&5.74&	0.71&0.63	&	6466	&	6466& 6361 & 6369	\\
00265	&	0.57375	&	12865	&	0.63272	& $-0.84$& $-0.91$	& $-1.10$&	$-1.28$	&5.28&5.32&5.69&5.78&	0.71&0.63	&	6466	&	6466& 6340 & 6348	\\
00268	&	0.47287	&	08672	&	0.53916	& $-0.63$& $-0.72$	& $-0.96$&	$-1.17$	&4.69&4.73&5.17&5.27&	0.80&0.68	&	6661	&	6661& 6549 & 6560\\
...&...&...&...&...&...&...&...&...&...&...&...&...&...&...&...&...&...\\
\hline

		\end{tabular}
\end{table*}

\section{Differences in the light curve parameters of the pairs}

Despite the identical-looking  light curves of the selected pairs, the differences between some of the Fourier parameters of the two stars are larger than $3\sqrt{{\sigma_{param_l}}^2+{\sigma_{param_s}}^2}$ in several cases. The errors of the Fourier parameters are calculated using the Monte-Carlo routine of Period04. 

\begin{figure}
\begin{center}
\includegraphics[width=.45\textwidth]{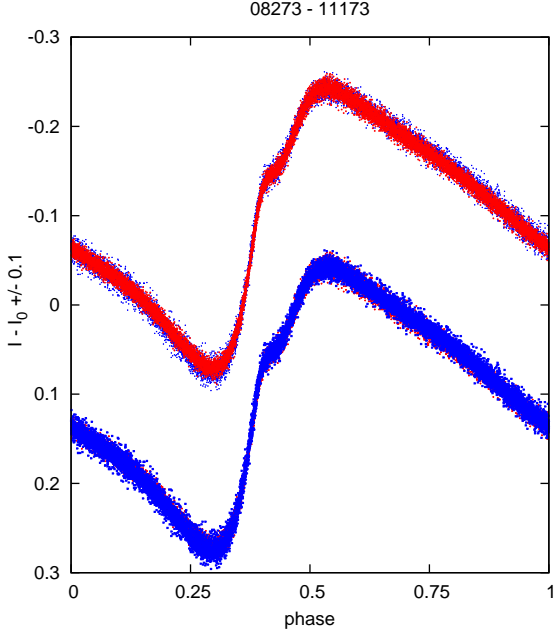}
\caption{An example for light curves looking to be identical but with Fourier-parameter differences significantly larger than their $3\sigma$ estimates. The two light curves are the $I$-band data of 08273 (red) and 11173 (blue) over-plotted on each other in different order. The period difference of this pair is 0.088 d.}
\label{fig:error}
\end{center}
\end{figure}

Fig.~\ref{fig:error} shows an example. The light curves of these stars seems to be absolutely equiform, however, the $a_1, R_{21}, \varphi_{21},  \varphi_{31}, \varphi_{41}$ parameter-differences  are as large as 15, 6, 9, 12, and 7 times of their estimated $\sigma$ uncertainty, respectively. 

We conclude that the calculated errors underestimate the true uncertainty of the parameters significantly, especially for the small-scatter, large data-number $I$-band data. 

Moreover, the low-order Fourier parameters do not reflect the differences in the fine details of the light curves (bump/hump features etc.) reliably.

This is why, in the course of the selection process, we relied on careful visual inspection of the light curves rather than selecting the pairs based on the errors
of the Fourier parameter differences.

Notwithstanding that the selected pairs are supposed to have no differences in shape, their Fourier parameters may not be completely identical. It is also checked whether there are any systematics in these small differences. Table~\ref{tab:diff} summarises the statistical properties of the low-order Fourier-parameter differences for the 149 pairs. The average $\sigma$ errors of the Fourier-parameter differences and their mean, median and s.dev. values  are given in the four lines of Table~\ref{tab:diff}.

The detected mean/median values of the parameter differences are close to their typical $\sigma$ errors, however, this very small, but systematic offsets are statistically significant with the exceptions of the total amplitude ($A_I$) and the $R_{31}$, $R_{41}$ amplitude ratios. These small offsets indicate that, in spite of the very careful visual inspection, the light curves of the members of the pairs are still not completely identical. However,  these differences  are actually below the detection limit in individual cases and are much smaller to have any influence on the calculated physical parameters of the stars.

\begin{table}
\label{tab:diff}
	\centering    
	\caption{Statistics of the $I$-band Fourier-parameter differences of the 149 pairs.}
\begin{tabular}{@{\hspace{0pt}}l@{\hspace{3pt}}r@{\hspace{3pt}}r@{\hspace{3pt}}r@{\hspace{3pt}}r@{\hspace{3pt}}r@{\hspace{3pt}}r@{\hspace{3pt}}r@{\hspace{3pt}}r@{\hspace{3pt}}r@{\hspace{3pt}}r}
\hline 
&${{A_I}}$&${{a_1}}$&${R_{21}}$&${R_{31}}$&${R_{41}}$&${R_{51}}$&$\phi_{21}$&$\phi_{31}$&$\phi_{41}$&$\phi_{51}$\\
\hline
$\overline{\sigma_\Delta}$&&.0005&.004&.004&.004&.004&.011&.019&.035&.063\\
mean$_\Delta$&   .000 & $-.001$ & .005 & .000 & .000 & .002 & $-.004$ & .016 & $-.032$ & $-.043$\\
median$_\Delta$& .000 & $-.001$ & .005 & .000 & .000 & .002 & $-.001$ & .017 & $-.020$ & $-.038$\\
s.d.$_\Delta$&   .012 &  .004 & .005 & .006 & .007 & .006 &  .021 & .027 &  .058 &  .112\\

\hline
		\end{tabular}
\end{table}

\section{Discussion / Conclusions}

The results documented in the previous sections have shown that the light-curve shape of RRab stars is not a unique function of the pulsation period, instead, RRab stars with quasi identical-shape light curves even in different wavelength bands do exist at different pulsation periods. 

It is a natural conclusion that this phenomenon may be connected to the Oosterhoff dichotomy and the period shift effect \citep{LeeC99,San04,Kun09}, at least in part. Although the Oo dichotomy has been suggested to occur simply as a consequence either of statistics or of systematics affecting previous investigations  based on homogeneous spectroscopic studies of large sample of field RRL star in the recent papers by \cite{Fabrizio2019,Fabrizio2021}, the two Oo groups can be clearly identified  in the Galactic bulge. Moreover, the spatial and  kinematical distributions and the mean physical properties ([Fe/H], $T_{\mathrm{eff}}$) of the two Oo populations are different in the bulge sample \citep{Prudil19a,Prudil19b}.

Most of the variables of the short-period sample are OoI-type variables while many of their long-period pairs belong to the OoII class according to their positions shown in Fig.~\ref{fig:pairs-plot}. 

The ratio of the mean differences between the log$P$ and the [Fe/H] values of the similar-shaped light-curve pairs is $-0.118$. This is in excellent agreement with the $-0.116$ period-shift value given by \cite{San82}.

OoI stars are less metal-poor than OoII stars, typically.  If the light curves of an OoI and an OoII star are very similar, then their Fourier amplitudes and phase-difference values have to be similar, too. Only their periods differ. Taking into account that the photometric metallicity formulae estimate the [Fe/H] as a function of $P$, $A$ and $\varphi_{31}$, in principle, it can happen that at different periods, the light curve parameters of different metallicity OoI and OoII RRL stars are identical.

However, to our best knowledge no detection of completely identical-shape light curves of OoI- and OoII-type stars has been ever shown.

\begin{figure}
\begin{center}
\includegraphics[width=.45\textwidth]{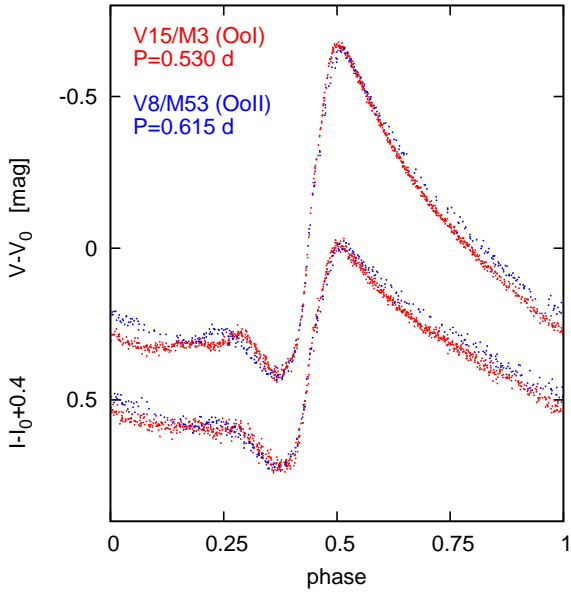}
\caption{Comparison of the $V$- and $I$-band light curves of similar amplitude stars in an OoI (V15/M3) and an OoII-type (V8/M53) globular clusters. The [Fe/H] of these clusters are $-1.46$ and $-2.00$, and the period difference between the two stars is 0.085 d. The amplitudes of the stars are the same in both photometric bands, however, their light-curve structure shows significant differences. }
\label{fig:m3comp}
\end{center}
\end{figure}

\begin{figure*}
\begin{center}
\includegraphics[width=1.0\textwidth]{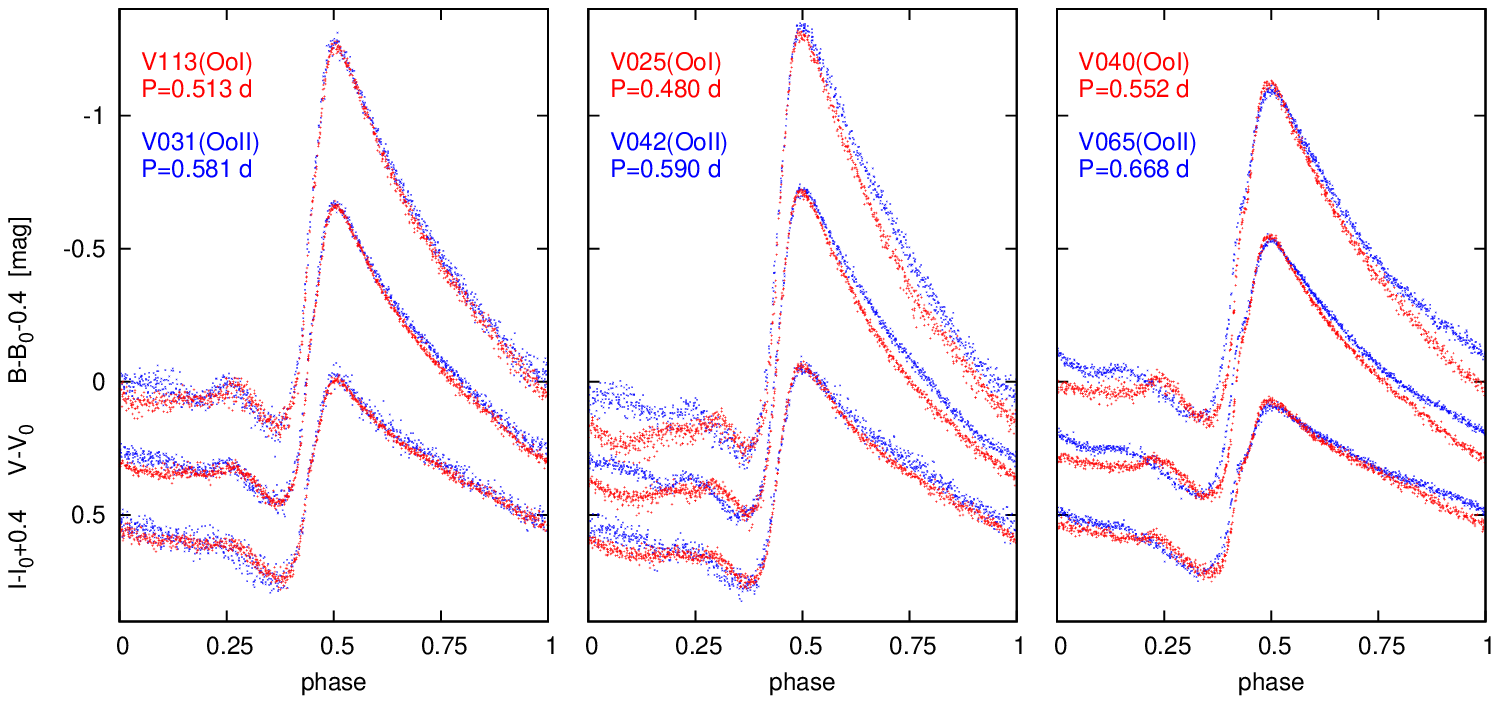}
\caption{Comparison of the light curves of Oo-type I and Oo-type II RRab stars in M3. The differences between the light-curve shapes of similar-amplitude stars at different periods are evident.}
\label{fig:m3}
\end{center}
\end{figure*}

The possible similarity of the light-curve shapes of OoI- and OoII-type RRL stars of similar amplitudes at different periods are checked using two examples of globular cluster variables. The light curves of variables in the M3 (OoI, [Fe/H]$=-1.46$) and the M53 (OoII, [Fe/H]=$-2.00$) globular clusters are compared, and also the light curves of some OoI and OoII stars in M3, despite the fact that the metallicities, in the latter case, are supposed to be the same. The data published in \cite{Jur17} and \cite{Bram11} are used for the M3 and M53 RRLs, respectively.
Figs.~\ref{fig:m3comp} and ~\ref{fig:m3} show the results. The period differences of the OoI and OoII stars compared in these figures are $0.07-0.11$ d, the amplitudes are the same both in the $V$ and in the $I$ bands, but significant differences in the light-curve shapes are evident in each plot.

These trial checks do not rule out the possibility that similar light-curve-shape OoI and OoII RRLs may occur in globular-cluster data but they show that the existence of these pairs cannot be explained simply by the Oo dichotomy.

Notwithstanding that the physical properties determined from the light-curve parameters in Sect.~\ref{sect:phys} hold some uncertainties, the detected mean differences between the physical properties of the long- and the short- period sequences are statistically significant on a 149-element sample.

The estimated physical parameters help to identify what type of RRLs are supposed to show identical-shape light curves at different pulsation periods.

Measuring these differences ($Param_\mathrm{l}-Param_\mathrm{s}$) as the function of their period differences we obtain the following connections: \newline
${\Delta{\mathrm{[Fe/H]}}}=-4.51{\Delta{P}}-0.02$, s.d. 0.03; \newline
${\Delta{\mathrm{[Fe/H]}}^*}=-6.33{\Delta{P}}-0.02$, s.d. 0.04; \newline
${\Delta{M_V}}=-1.54{\Delta{P}}+0.27P_{\mathrm{l}}-0.16$, s.d. 0.01;  \newline
${\Delta{R/R_\odot}}=6.90{\Delta{P}}+0.02$, s.d. 0.02; \newline
${\Delta{R/R_\odot}^*}=7.91{\Delta{P}}+0.02$, s.d. 0.02; \newline
${\Delta{T_{\mathrm{eff}}}=-1851\Delta{P}}$, s.d. 34; \newline
${\Delta{T_{\mathrm{eff}}^*}}=-1766{\Delta{P}}$, s.d. 35.

The physical parameter differences of the similar-amplitude OoI and OoII globular-cluster variables shown in Figs.~\ref{fig:m3comp} and ~\ref{fig:m3} do not fit these relations. Supposing that the M3 globular cluster is mono-metallic, no [Fe/H] difference between the OoI and the OoII-type variables of the cluster is probable.
In the case of the other example, notwithstanding that the estimated $M_V$ and $R$ differences of the compared M3 and M53 variables fit these relations, the differences between their [Fe/H] and ${T_{\mathrm{eff}}}$ values are twice as large as the relations given above would indicate.

However, even if these parameter-difference relations are met, we cannot be sure that there is no hidden extra parameter that accounts for the identical-shape light curves of these stars. 

Summarising our findings, we identified pairs of RRab stars with different pulsation periods and different physical properties but with identical-shape light curves according to all the available photometric observations. To show such identical-shape light variations the complete dynamical structures of these stars have to be very similar.  However, the different metallicity, and consequently the opacities of the atmosphere of these stars make it very unlikely. The shocks, forming the shapes of the bump, and the hump features of the light curves are not expected to behave so similarly in RRLs with noticeably different fundamental parameters, either. Quoting \cite{Bono20}: "The change in the physical structure also implies variation in the pulsation properties." 
Therefore, to explain the light curve similarity of these pairs is a challenging task for modelling RRL pulsation.

The OGLE $I$-band data yield very accurate light curves but the $V$-band data are sparse, and the VVV and the MACHO light curves have large scatter and some amplitude ambiguity. Therefore, time-series observations in different bands of some of the best candidate twins would be essential to strengthen the results published in this paper. Observations at shorter wavelengths were especially very informative, however, taking into account the large reddening of the bulge stars, there are limitations in obtaining these data.

\section{Acknowledgement}
The authors thank Gergely Hajdu for valuable comments and for providing us the corrected VVV photometry of the studied variables.
The research leading to these results has been supported by the Hungarian National Research, Development and Innovation Office (NKFIH) grants NN-129075 and K-129249. We gratefully thank the work of the OGLE team, this paper is based primarily on the OGLE-BULGE-RRL data. We also acknowledge the data taken with the VISTA telescope from the ESO Public Survey program ID 179.B-2002. This paper utilises public domain data obtained by the MACHO Project, jointly funded by the US Department of Energy through the University of California, Lawrence Livermore National Laboratory under contract No. W-7405-Eng-48, by the National Science Foundation through the Center for Particle Astrophysics of the University of California under cooperative agreement AST-8809616, and by the Mount Stromlo and Siding Spring Observatory, part of the Australian National University.

\section{Data Availibility Statment}
The paper utilises photometric data of the  OGLE-IV, the VVV and the MACHO surveys, which are publicly available.

\bibliographystyle{mnras}
\bibliography{Jurcsik}

\bsp	
\label{lastpage}
\end{document}